\documentclass{article}
\usepackage{spconf,amsmath,graphicx,url,times}
\usepackage{color}
\usepackage[nolist]{acronym}
\usepackage{mathtools}
\usepackage{cite}  
\usepackage{caption}
\usepackage{subcaption}
\usepackage{wrapfig}
\usepackage{environ}         
\usepackage{etoolbox}  
\usepackage{hyperref}
\usepackage{float}
\usepackage{amssymb}
\usepackage{etoolbox}
\usepackage{glossaries}

\newcommand\norm[1]{\lVert#1\rVert}

\title{BINAURAL SPEECH ENHANCEMENT USING COMPLEX CONVOLUTIONAL RECURRENT NETWORKS}

 \name{Vikas Tokala$^{^1}$,
      Eric Grinstein$^{1}$,
      Mike Brookes$^{^1}$,
      Simon Doclo$^{^2}$,
      Jesper Jensen${^{3,4}}$,
      Patrick A. Naylor $^1$\sthanks{This work was supported by funding from the European Union’s Horizon 2020 research and innovation programme under the Marie Skłodowska-Curie grant agreement No 956369 and the UK Engineering and Physical Sciences Research Council [grant number EP/S035842/1]}} 
\address{$^{1}$Department of Electrical and Electronic Engineering, Imperial College London, UK\\
         $^{2}$ Department of Medical Physics and Acoustics, University of Oldenburg, Germany.\\
        $^{3}$  Demant A/S, Smørum, Denmark.\\
        $^{4}$ Department of Electronic Systems, Aalborg University, Denmark
}
\begin{acronym}
\acro{SNR}{Signal-to-Noise Ratio}
\acro{fwSegSNR}{frequency-weighted Segmental SNR}
\acro{SOBM}{STOI-optimal Binary Mask}
\acro{STOI}{Short-Time Objective Intelligibility}
\acro{WSTOI}{Weighted STOI}
\acro{HSWOBM}{High-resolution Stochastic WSTOI-optimal Binary Mask}
\acro{STFT}{Short Time Fourier Transform}
\acro{ILD}{Interaural Level Differences}
\acro{IPD}{Interaural Phase Differences}
\acro{ITD}{Interaural Time Differences}
\acro{LSA}{Log Spectral Amplitude}
\acro{OM-LSA}{Optimally-modified Log Spectral Amplitude}
\acro{DNN}{Deep Neural Network}
\acro{RNN}{Recurrent Neural Network}
\acro{ERB}{Equivalent Rectangular Bandwidth}
\acro{DFT}{Discrete Fourier Transform}
\acro{VSSNR}{Voiced-Speech-plus-Noise to Noise Ratio}
\acro{SPP}{Speech Presence Probability}
\acro{MBSTOI}{Modified Binaural STOI}
\acro{ReLU}{Rectified Linear Unit}
\acro{SNR}{Signal-to-Noise Ratio}
\acro{ISTFT}{Inverse STFT}
\acro{TF}{Time-Frequency}
\acro{WGN}{White Gaussian Noise}
\acro{RTF}{Relative Transfer Function}
\acro{EVD}{Eigen Value Decomposition}
\acro{PSD}{Power Spectral Density}
\acro{IRM}{Ideal Ratio Mask}
\acro{NPM}{Normalized Projection Misalignment}
\acro{CRM}{Complex Ratio Mask}
\acro{FTB}{Frequency Transformation Block}
\acro{FTM}{Frequency Transformation Matrix}
\acro{FAL}{Frequency Attention Layer}
\acro{MWF}{Multichannel Wiener Filter}
\acro{MVDR}{Minimum Variance Distortionless Response}
\acro{MSC}{Magnitude Squared Coherence}
\acro{VAD}{Voice Activity Detector}
\acro{PESQ}{Preceptual Evalution of Speech Quality}
\acro{MLP}{Multi-Layer Perceptrons}
\acro{AED}{Auto Encoder-Decoder}
\acro{CED}{Convolutional Encoder-Decoder}
\acro{CNN}{Convolutional Neural Network}
\acro{CRN}{Convolutional Recurrrent Network}
\acro{LSTM}{Long short-term memory}
\acro{TNN}{Transformer Neural Networks}
\acro{NLP}{Natural Language Processing}
\acro{IBM}{Ideal Binary Mask}
\acro{IRM}{Ideal Ratio Mask}
\acro{SI-SNR}{Scale Invariant SNR}
\acro{GRU}{Gated Recurrent Unit}
\acro{HRIRs}{Head Related Impulse Responses}
\acro{SSN}{Speech Shaped Noise}
\acro{HATS}{Head and Torso Simulator}
\acro{PReLU}{Parametric Rectified Linear Unit}
\acro{BSOBM}{Binaural STOI-Optimal Masking}
\acro{BMWF}{Binaural MWF}
\acro{SegSNR}{Segmental SNR}
\acro{RIR}{Room Impulse Responses}
\acro{BiTasNet}{Binaural TasNet}
\acro{SI-SNR}{Scale Invariant SNR}
\acro{BCCRN}{Binaural Complex Convolutional Recurrent Network}
\acro{BRIR}{Binaural Room Impulse Responses}

\end{acronym}
\begin{document}

\ninept
\maketitle

\begin{sloppy}

\begin{abstract}
 From hearing aids to augmented and virtual reality devices, binaural speech enhancement algorithms have been established as state-of-the-art techniques to improve speech intelligibility and listening comfort. In this paper, we present an end-to-end binaural speech enhancement method using a complex recurrent convolutional network with an encoder-decoder architecture and a complex LSTM recurrent block placed between the encoder and decoder. A loss function that focuses on the preservation of spatial information in addition to speech intelligibility improvement and noise reduction is introduced. The network estimates individual complex ratio masks for the left and right-ear channels of a binaural hearing device in the time-frequency domain. We show that, compared to other baseline algorithms, the proposed method significantly improves the estimated speech intelligibility and reduces the noise while preserving the spatial information of the binaural signals in acoustic situations with a single target speaker and isotropic noise of various types.
\end{abstract}


\section{Introduction}
\label{sec:intro}
Binaural speech enhancement has gained significant attention as a state-of-the-art approach for enhancing speech in various applications, including hearing aids and augmented/virtual reality devices \cite{Doclo2008, Guiraud2022}. Binaural signals preserve the spatial characteristics of sounds which help listeners in noisy acoustic environments achieve better speech intelligibility and accurate sound source localization \cite{Hawley2004}. The fundamental spatial cues that help in localizing sounds and improving intelligibility are \ac{ILD} and \ac{IPD} or \ac{ITD} \cite{Blauert1997}. Previously proposed methods for binaural speech enhancement include multichannel Wiener filters \cite{Hadad2015, Klasen2006}, beamforming-based \cite{Doclo2008}, and mask-based enhancement methods \cite{Tokala2022, Moore2018b}. Binaural speech separation using time-domain \ac{CED} was proposed in \cite{Han2020}.  Recent advancements in deep learning techniques have led to remarkable improvements in monaural speech enhancement. These methods, whether applied in the time domain \cite{Stoller2018a, Luo2018} or the \ac{TF} domain \cite{Williamson2015,Tan2018, Yin2020a}, demonstrate impressive results.

Spectrograms are commonly used as input to networks in the \ac{TF} domain for speech enhancement \cite{Tokala2022, Yin2020a, Hu2020}. Many methods in this domain focus on enhancing only the magnitude of the spectrogram while using the noisy phase information for reconstructing the enhanced speech signal \cite{Tokala2022, Kim2020a}. To address optimal phase estimation and signal reconstruction, some approaches jointly estimate both magnitude and phase by utilizing complex-valued spectrograms. These methods have shown promising results and can outperform real-valued networks in monaural speech enhancement \cite{Kim2020a, Hu2020}.

The \ac{CRN} introduced in \cite{Tan2018} employed a \acf{CED} architecture with \ac{LSTM} blocks placed in between the encoder and decoder. In \cite{Hu2020}, a deep complex \ac{CRN} was trained to optimize the \ac{SI-SNR} for monaural speech signals. However, using a similar approach for binaural signals could be damaging to the interaural cues. More specifically, for the case of binaural signals, phase information is vital for preserving the \ac{IPD} values and the enhanced signals should preserve level differences as the noisy signal to retain \ac{ILD}. While the model may effectively reduce noise and enhance speech intelligibility, modifying the level and phase information could potentially impact the spatial information of the target, leading to a compromised ability for localization and spatial awareness \cite{Blauert1997, Beutelmann2006}. In \cite{Tokala2024}, a complex convolutional attention-based transformer network has been proposed which uses a similar loss function based on interaural cues. 

Our proposed method, \ac{BCCRN}, uses a complex-valued \ac{CED}-based recurrent network for binaural speech enhancement and introduces terms in the loss function to improve speech intelligibility while preserving the interaural cues based on human perception of the target speech signal with a smaller complex recurrent network compared to \cite{Tokala2024}.  


\section{\ac{BCCRN} Model Architecture}
\label{sec:method}
 \begin{figure*}
    \centering
    \includegraphics[width=\textwidth]{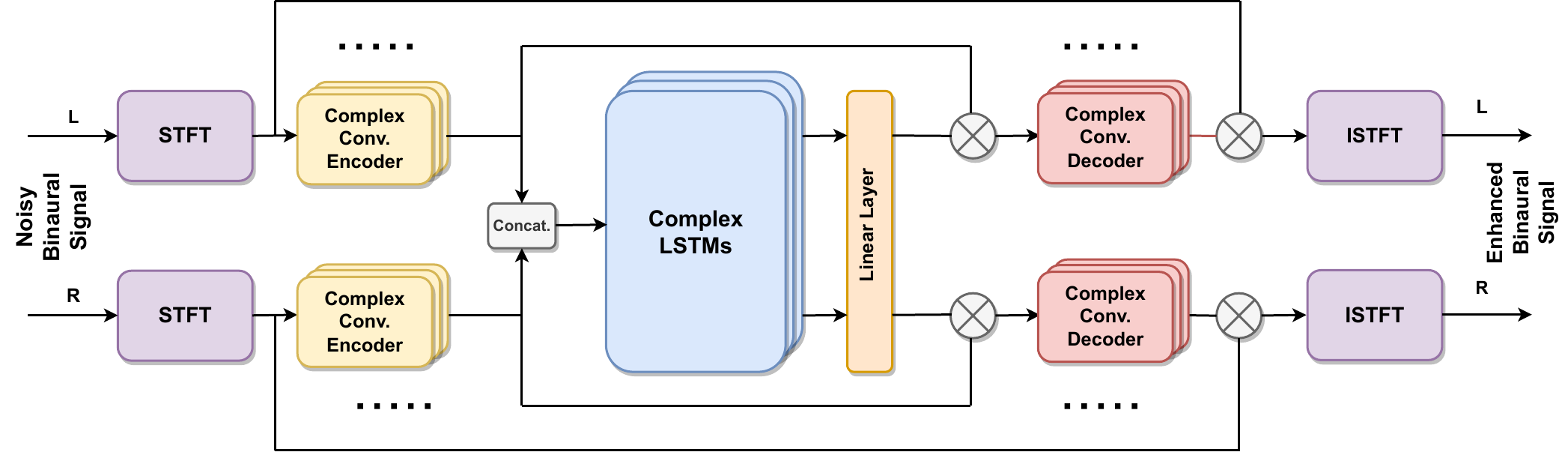}
     \vspace{-0.2cm}
    \caption{Model architecture of \acf{BCCRN}.}
    \label{fig:BlockDiag}
     \vspace{-0.2cm}
\end{figure*}

The proposed \ac{BCCRN} is trained to estimate individual \ac{CRM} for each channel.  The block diagram of the model architecture is shown in Fig.~\ref{fig:BlockDiag}. The \ac{STFT} blocks transform the input signal into the \ac{TF} domain. The encoder block is made of 6 complex convolutional layers with \ac{PReLU} activation and employs batch normalization. Separate encoder and decoder blocks are used for the left and right-ear channels to estimate the individual \ac{CRM}s. The decoder block consists of 6 complex convolutional layers which are symmetric in design to the layers in the encoder to reconstruct the signal. The encoder extracts high-level features from the input spectrograms and reduces the resolution of the input. The decoder rebuilds the low-resolution features back to the original size. The encoded information from the left and right encoder blocks is concatenated and provided to the complex LSTM block. The \ac{LSTM} layers are designed to model the frequency dependencies. Skip connections are placed between the encoder and decoder layers based on the \ac{CRN} architecture which improves the information flow and facilitates network optimization \cite{Tan2018}. The individual decoders output individual \ac{CRM}s that are applied to the left and right channels of the noisy binaural signal for enhancement. The \ac{ISTFT} blocks transform the signal back into the time domain.

\section{Signal Model and Loss Function}
\label{sec:sigmod}

The noisy time-domain input signal for the right channel, $y_{R}$, is given by 
\begin{equation}
    y_{R}(t) = s_{R}(t) + v_{R}(t),
\end{equation}
where $s_{R}$ is the anechoic clean speech signal, $v_{R}$ is the noise and  $t$ is the time index. The \ac{STFT} is used to transform the signals into the \ac{TF} domain and the respective \ac{TF} representations are $Y_{R}(k,\ell)$, $S_{R}(k,\ell)$ and $V_{R}(k,\ell)$ with $k$ and $\ell$ being the frequency and time frame indices respectively. The left channel is described similarly with an $L$ subscript. For brevity, the $L$ and $R$ indices are omitted from the remainder of this paper. The estimated \ac{CRM}, $M(k,\ell)$ is applied to the noisy signal $Y(k,\ell)$ to obtain the enhanced speech signal $\hat{S}$. Individual channels are enhanced by applying the estimated \ac{CRM}, $\left({M}_r + j {M }_i \right)$ to the complex-valued noisy signal $\left({Y}_r + j {Y }_i\right)$ in the \ac{TF} domain (omitting $k$ and $\ell$ indices),

\begin{equation}
    \hat{S}_r + j \hat{S}_i = \left({M}_r + j {M }_i \right) \cdot \left({Y}_r + j {Y }_i\right),
\end{equation}
where $r$ and $i$ indicate the real and imaginary parts. The computed \ac{CRM} is given by 
\begin{equation}
    {M}_r + j {M }_i = \frac{\hat{S}_r + j \hat{S}_i}{{Y}_r + j {Y }_i} = \frac{{Y}_r \hat{S}_r+{Y}_i \hat{S}_i}{Y_r^2 + Y_i^2} + j \frac{{Y}_r \hat{S}_i-{Y}_i \hat{S}_r}{Y_r^2 + Y_i^2}.
\end{equation}

\subsection{Loss Function}\label{sec:LossFunc}
The loss function for model training consists of four terms and optimizes the network for noise reduction, intelligibility improvement, and interaural cue preservation. The loss function $\mathcal{L}$ is given by
\begin{equation} \label{lossfunc}
\mathcal{L} = \alpha \mathcal{L}_{SNR} + \beta \mathcal{L}_{STOI} +   \gamma \mathcal{L}_{ILD} + \kappa \mathcal{L}_{IPD},
\end{equation}
where $\mathcal{L}_{SNR}$ is the \ac{SNR} loss, $\mathcal{L}_{STOI}$ is the \ac{STOI} \cite{Taal2010} loss, and $\mathcal{L}_{ILD}$ and $\mathcal{L}_{IPD}$ are the proposed \ac{ILD} and \ac{IPD} error losses which are functions of both $\hat{S}_L$ and $\hat{S}_R$. The parameters $\alpha$, $\beta$, $\gamma$, and $\kappa$ are the weights applied to each term. 

$\mathcal{L}_{SNR}$ is defined as the mean of the left and right-ear channel values and append a negative sign to maximize the \ac{SNR} value, such that $\mathcal{L}_{SNR} = - \left(\ac{SNR}_{L} + \ac{SNR}_{R} \right)/2$. 
The \ac{SNR} of the enhanced signal, $\hat{\mathbf{s}}$, is defined as
\begin{equation} \label{snr}
   \ac{SNR}( \mathbf{s},\hat{\mathbf{s}}) =  10 \log_{10} \left( \frac{\norm{\mathbf{s}}^2}{\norm{\mathbf{e}_{noise}}^2}\right),
\end{equation}
where $\mathbf{e}_{noise} = \hat{\mathbf{s}} - \mathbf{s}$ with $\mathbf{s}$ and $\hat{\mathbf{s}}$ being the clean and enhanced signal vectors respectively, and $\norm{.}$ is the L2 norm.  

To optimize the intelligibility of the enhanced speech signals, \ac{STOI} is computed for left and right channels individually and averaged, and a negative sign is appended to maximize the \ac{STOI} so that $\mathcal{L}_{STOI}~=~ - \left(\ac{STOI}_{L} + \ac{STOI}_{R} \right) /2$\cite{Manuel2023}. Including $\mathcal{L}_{STOI}$ helps the network optimize individual \ac{CRM}s to maximize intelligibility in the enhanced signals.

To ensure the preservation of interaural cues in the enhanced binaural speech using two separate \ac{CRM}s, minimizing the \ac{ILD} and \ac{IPD} errors of the target speech is enforced using the loss function. The \ac{ILD} and \ac{IPD} for the clean speech signal are given by

\begin{equation} \label{ild}
    ILD_S(k,\ell) = 20 \log_{10} \left ( \frac{\vert S_{L}(k,\ell) \vert }{ \vert S_{R}(k,\ell) \vert}\right),
\end{equation}
\begin{equation}
\label{ipd}
    IPD_S (k, \ell) = \arctan \left ( \frac{S_{L}(k,\ell) }{S_{R}(k,\ell)} \right).
\end{equation}

\noindent The \ac{ILD} and \ac{IPD} for the enhanced speech, $\hat{S}$, are calculated similarly. In order to restrict the \ac{ILD} and \ac{IPD} errors to the speech-active regions, an \ac{IBM} \cite{Wang2005b} $\mathcal{M}$ is computed by selecting the \ac{TF} bins which have energy above a threshold. The energy $E(k,\ell)$ of the clean signal is given by

\begin{equation}
    E(k,\ell) = 10 \log_{10} {\vert {S}(k,\ell) \vert}^2.
\end{equation}
 The \ac{IBM} $\mathcal{M} (k,\ell)$ that defines the speech active \ac{TF} tiles is then defined as,
\vspace{-0.15cm}
\begin{equation}
    \mathcal{M}(k,\ell) = \begin{dcases}
    1 & E(k,\ell) > \max_{\ell} \left ( E(k,\ell) \right) - \mathcal{T} \\
    0 & \mathrm{otherwise}.
    \end{dcases}
\end{equation}
where $\max_l \left ( E(k,\ell) \right) $ is the maximum energy computed for each frequency bin, $k$. Individual \ac{IBM}s, $\mathcal{M}_L$ and $\mathcal{M}_R$ are computed for the left and right-ear channels. The final mask $\mathcal{M}$ is obtained by choosing the bins that have energy above the threshold, $\max_{\ell} \left ( E(k,\ell) \right) - \mathcal{T}$, in both channels and is given by

\begin{equation}
    \mathcal{M}(k,\ell) = \mathcal{M}_L(k,\ell) \odot \mathcal{M}_R (k,\ell),
\end{equation}
\noindent where $\odot$ denotes the Hadamard product. For training and evaluation $\mathcal{T}=20$~dB was used \cite{Wang2005b}. The human auditory system interprets \ac{ILD}s and \ac{IPD}s differently based on the frequency range. Human spatial hearing relies primarily on interaural phase difference cues for frequencies below 1500 Hz, while interaural level difference cues play a crucial role for frequencies above 1500 Hz \cite{Blauert1997}. From the estimated \ac{IBM} $\mathcal{M}(k,\ell)$, separate masks for \ac{ILD} and \ac{IPD} cues are computed based on human spatial hearing. For choosing the \ac{IPD} cues in the speech active bins below $f_p =1500$~Hz, $\mathcal{M}(k,\ell)$ for $ k~\leqslant~K_p$ is selected where $K_p = f_p \times N_{fft}/f_s$ with $N_{fft}$ and $f_s$ being the FFT length and sampling frequency respectively. Similarly, for choosing the \ac{ILD} cues $\mathcal{M}(k,\ell)$ for $k~>~K_p$ is selected.
The $\mathcal{L}_{ILD}$ and $\mathcal{L}_{IPD}$ terms are given by
\begin{equation} \label{ildloss}
  \mathcal{L}_{ILD} = \frac{1}{N_{ld}} \sum_{k > K_p, \ell} \mathcal{M}(k,\ell)  \left(  \vert ILD_{S} (k,\ell) - ILD_{\hat{S}} (k,\ell) \vert \right ),
\end{equation}
\begin{equation}\label{ipdloss}
    \mathcal{L}_{IPD} = \frac{1}{N_{pd}} \sum_{k \leqslant K_p, \ell} \mathcal{M}(k,\ell) \vert IPD_{S} (k,\ell) - IPD_{\hat{S}} (k,\ell) \vert
\end{equation}

\noindent where $N_{ld} = \sum_{k > K_p,\ell} \mathcal{M}(k,\ell) $  and $N_{pd} = \sum_{k \leqslant K_p,\ell} \mathcal{M}(k,\ell) $  are the total number of speech-active frequency and time bins determined from the mask for \ac{ILD} and \ac{IPD} cues respectively. To preserve the interaural cues of the target speaker, the network optimization is guided by using masks based on the target speech. The errors in \ac{ILD} and \ac{IPD} are calculated in the time-frequency domain, while losses related to \ac{SNR} and \ac{STOI} are computed in the time domain through waveform synthesis using the \ac{ISTFT}.

\section{Experiments}
\label{sec:Experiments}

\subsection{Datasets}
To generate binaural speech data, monaural clean speech signals were acquired from the CSTR VCTK corpus \cite{Yamagishi2019} and then spatialized using \ac{HRIRs} from \cite{Kayser2009}. The selected speech corpus contains approximately 13 hours of spoken English data recorded by 110 speakers with diverse accents. From this dataset, each two-second utterance was converted into binaural form with distinct left and right-ear channels. The dataset consisted of 20,000 speech utterances that were divided into training, validation, and testing sets. Diffuse isotropic noise was generated using noise signals from the NOISEX~-~92 database \cite{Varga1993}. Uncorrelated noise sources were evenly placed at intervals of $5^\circ$ in the azimuthal plane to create isotropic noise\cite{Moore2018b} using \ac{HRIRs} from \cite{Kayser2009}. For generating binaural signals, the target speech was placed randomly within the frontal plane ($-90^\circ$ to $+90^\circ$), utilizing \ac{HRIRs} recorded with a HATS \cite{Kayser2009}. For training, isotropic noise was added to the VCTK corpus\cite{Yamagishi2019} so that $(SNR_L + SNR_R)/2$ lies between -7~dB and 16~dB. The noise types used for training are \ac{WGN}, \ac{SSN}, factory noise, and office noise. An unseen engine noise type was included in the evaluation set. The datasets were generated in the anechoic condition. The evaluation set comprises speech signals from both the VCTK corpus \cite{Yamagishi2019} and the TIMIT corpus \cite{Garofolo1993}. In this set, random target azimuths in the frontal azimuthal plane are assigned, and isotropic noise is introduced at varying \ac{SNR}s ranging from -6 dB to 15 dB. The speaker was positioned at a distance of either 0.8~m or 3~m randomly at a fixed elevation of $0^\circ$.

\subsection{Training setup}
To compute the \ac{STFT}, an FFT length of 512, a window length of 25~ms, and a hop length of 6.25~ms were employed. A sampling rate of 16~kHz was utilized for all signals.

 \noindent\textbf{\acf{BCCRN}}: The number of channels used in the model's convolutional layers for the complex-valued encoder and decoder block layers are $\{32, 64, 128, 256, 256, 256\}$, with a stride of 2 in the frequency and 1 in the time dimension with a kernel size of (5,1) and all the convolutions in these layers are causal. 8 layers of bidirectional complex-valued LSTMs with a hidden size of 128 was used. The model was implemented with Pytorch which provides native complex data support for most of the functions. The linear layer placed after the recurrent block has an input and output feature size of 1024. The Pytorch model was trained using the Adam optimizer, an initial learning rate of 0.001, and a multi-step learning rate scheduler to modify the learning rate with the validation loss. The model has around 5.7~million parameters and was trained for 100 epochs with an additional early stopping condition of no improvement in the validation loss for three consecutive epochs. The weights for the loss function $\alpha, \beta, \gamma, \kappa$ in \eqref{lossfunc} were assigned as \{1,10,1,10\} respectively. These weight values were selected to standardize the units of each individual loss function term. The terms involving SNR and ILD are computed in dB, IPD is calculated in radians and STOI is a bounded score ranging from 0 to 1. The model was trained with the proposed loss function described in \eqref{lossfunc}, named \ac{BCCRN}-SILP,  and for comparison, the model was trained to maximize the \ac{SNR}, named \ac{BCCRN}-S from \eqref{snr}.
\subsection{Baselines}
\textbf{\ac{BSOBM}}: A binaural speech enhancement method using \ac{STOI}-optimal masks proposed in \cite{Tokala2022}. Here a feed-forward \ac{DNN} was trained to estimate a \ac{STOI}-optimal continuous-valued mask to enhance binaural signals using dynamically programmed \ac{HSWOBM} as the training target \cite{Tokala2022}. To preserve the \ac{ILD}s, a better-ear mask was computed by choosing the maximum of the two masks.  The mask is used to supply \ac{SPP} to an \ac{OM-LSA} enhancer. The model was trained and evaluated on the same dataset as the proposed model.

\textbf{\ac{BiTasNet}}: A time-domain \ac{CED}-based network for binaural speech separation which was introduced in \cite{Han2020}. The best-performing version of the model, the parallel encoder with mask and sum, was modified and retrained for single-speaker binaural speech enhancement. The network was trained to maximize \ac{SNR}\cite{Han2020}. The encoder and decoders in the model had a size of 128, a feature dimension of 128, kernel size of 3 and 12 layers. All other parameters were adapted from the original article and the model has a size of 7 million parameters. The model was trained and evaluated on the same dataset used for the proposed method.

\section{Results and Discussion}
\label{sec:results}
\begin{figure*}[!h]
     \centering
     \begin{subfigure}[b]{0.33\textwidth}
         \centering
         \includegraphics[width=0.95\textwidth]{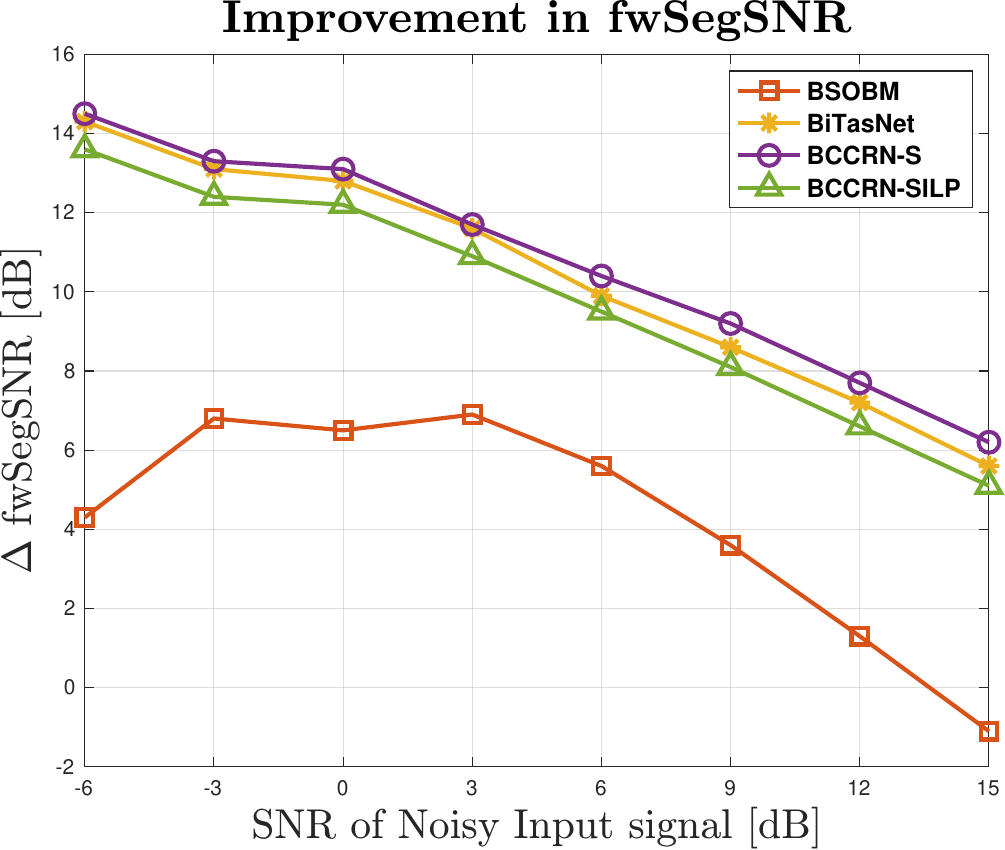} 
         \caption{}
         \label{fig:del_segsnr}
     \end{subfigure}
     \hfill
     \begin{subfigure}[b]{0.33\textwidth}
         \centering
         \includegraphics[width=0.95\textwidth]{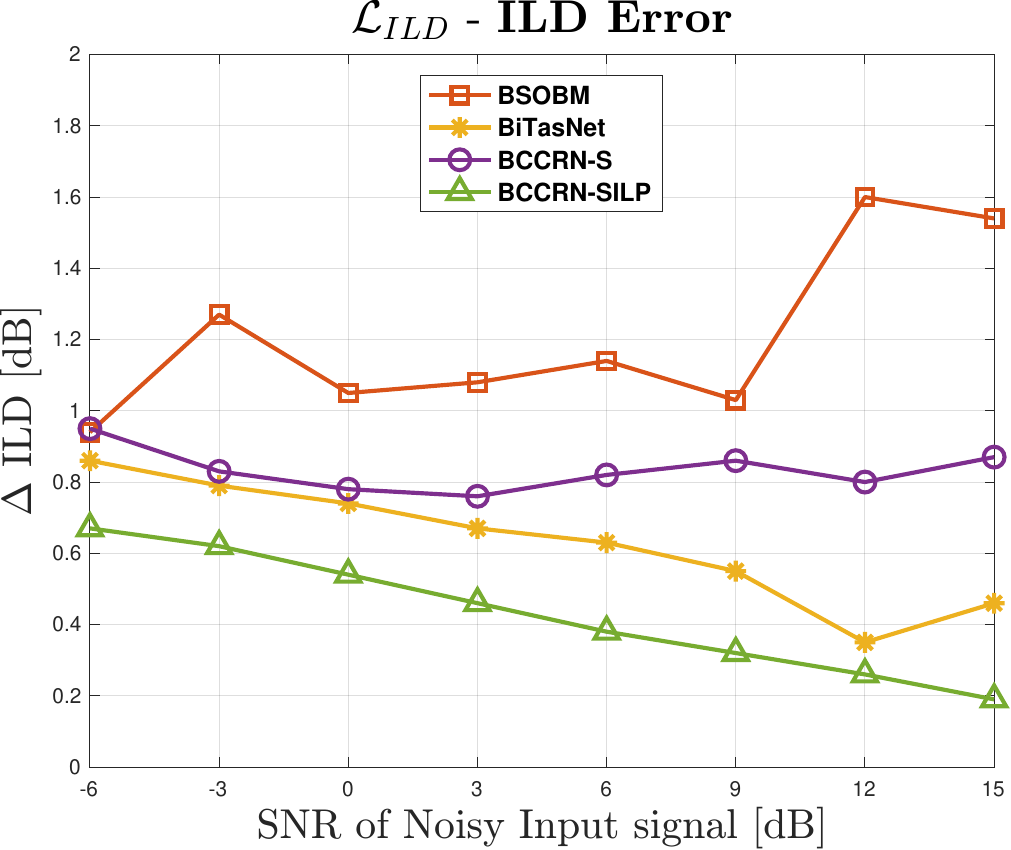}
         \caption{}
         \label{fig:del_ild}
     \end{subfigure}
     \hfill
     \begin{subfigure}[b]{0.33\textwidth}
         \centering
         \includegraphics[width=0.95\textwidth]{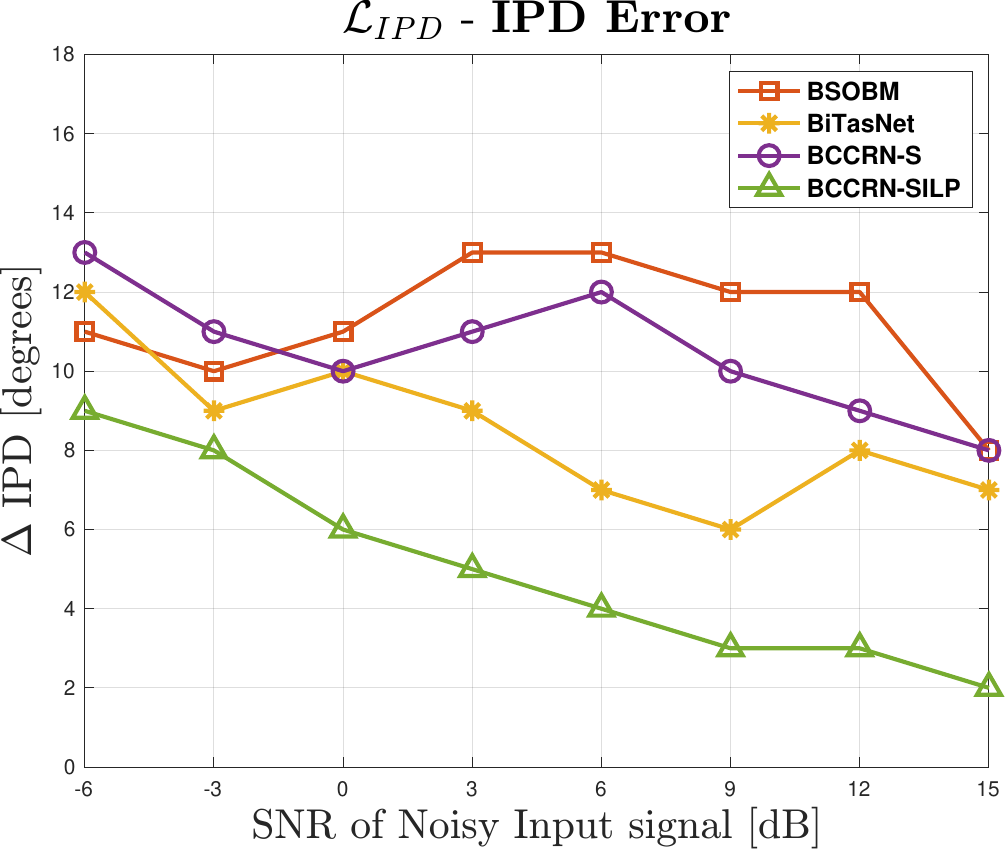}
         \caption{}
         \label{fig:del_ipd}
     \end{subfigure}
      \vspace{-0.3cm}
        \caption{Comparison of (a) improvement in \ac{fwSegSNR} (b) \ac{ILD} error \eqref{ildloss} and (c) \ac{IPD} error \eqref{ipdloss} for speech signals with isotropic noise averaged over all frames, frequency bins, and utterances.}
        \label{fig:noises}
\end{figure*}
\begin{figure}[!h]
    
         \includegraphics[width=0.45\textwidth]{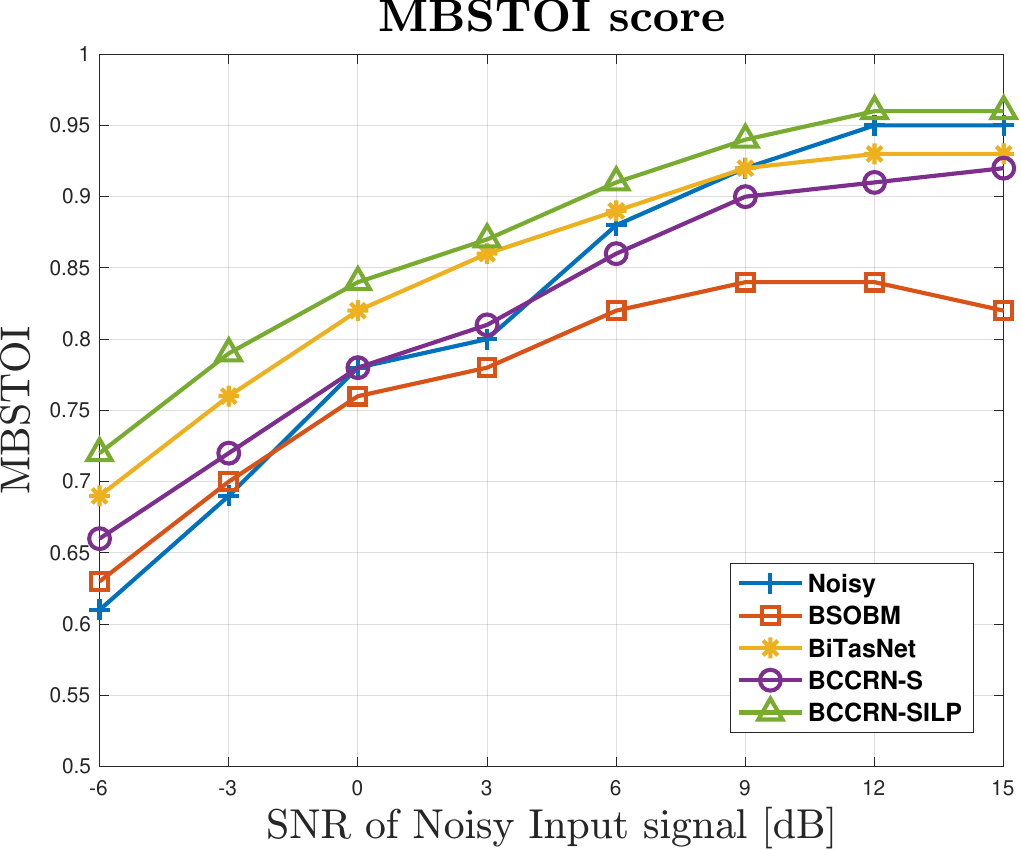}

        \caption{Comparison of the MBSTOI score for speech signals with isotropic noise after enhancement averaged over all utterances.  } 
        \vspace{-0.5cm} \label{methodsfig}
\end{figure}

The performance of the methods was assessed by evaluating 750 utterances from both VCTK \cite{Yamagishi2019} and TIMIT \cite{Garofolo1993} datasets for each of the 8 input \ac{SNR}s. The noise reduction capability of the methods was demonstrated through improvement in \ac{fwSegSNR}\cite{Brookes1997}. Objective binaural speech intelligibility of the enhanced signals was measured using the \ac{MBSTOI}\cite{Andersen2018} metric. The preservation of interaural cues was evaluated by calculating the error in \ac{ILD} and \ac{IPD} after processing, using equations \eqref{ildloss} and \eqref{ipdloss}, respectively. Figure~\ref{fig:del_segsnr} shows the improvement in \ac{fwSegSNR} \cite{Brookes1997} for different input \ac{SNR}s. \ac{BCCRN}-S has the best performance for almost all \ac{SNR}s, with noise reduction measured by the improvement in the \ac{fwSegSNR}, while \ac{BSOBM} has the lowest improvement. Nevertheless, the proposed model exhibits similar effectiveness to the \ac{BiTasNet} and \ac{BCCRN}-S in reducing noise. The model has better performance when trained to optimize the \ac{SNR} compared to the proposed loss function and provides an additional 1~dB of improvement in \ac{fwSegSNR} on average. A maximum improvement of about 14~dB \ac{fwSegSNR} is observed in the noisy input \ac{SNR}s and the amount of improvement observed tends to decrease as the input \ac{SNR} of the noisy signal improves for both the \ac{BCCRN} versions. Figures~\ref{fig:del_ild} and \ref{fig:del_ipd} show the \ac{ILD} and \ac{IPD} errors after enhancement computed using \eqref{ildloss} and \eqref{ipdloss}. The proposed model and loss function have the lowest error for both cues. The suggested model utilizing the \ac{SNR} loss function demonstrates comparable performance to the proposed loss function in reducing noise, but it does not prioritize preserving the interaural differences. The inclusion of additional terms in the loss function aids the network in better maintaining interaural differences. The observed \ac{ILD} error was under 1~dB and the \ac{IPD} error was under $10^\circ$ for all input \ac{SNR}s of the noisy signal. Also, the \ac{ILD} and \ac{IPD} errors for the proposed method tend to decrease with increasing input \ac{SNR} while this is not observed in the model with \ac{SNR} loss function and other methods. Figure \ref{methodsfig} shows the \ac{MBSTOI} of the enhanced signals. The proposed loss function had the best performance for all \ac{SNR}s and provided an average of 0.15 to 0.25 improvement in the \ac{MBSTOI} score. The \ac{BCCRN}-S has a lower intelligibility score even though it has the best noise reduction performance. 
Despite its improved ability to reduce noise, the \ac{BiTasNet} model demonstrates a lower binaural intelligibility score as measured by \ac{MBSTOI} shown in Fig.~\ref{methodsfig}. Informal listening tests revealed that the \ac{BiTasNet} produced more artefacts and reduced intelligibility of the speech.  Similar to \ac{fwSegSNR} and the error in interaural cues, \ac{BSOBM} has the lowest \ac{MBSTOI} score for enhanced signals. A common trend observed in binaural speech enhancement methods is the degradation of the \ac{MBSTOI} score at high input \ac{SNR}s due to processing \cite{Tokala2022} as the input signals inherently have a higher \ac{MBSTOI} score. However, the proposed loss function does not deteriorate the \ac{MBSTOI} score, preserves the intelligibility, and provides an improvement at all \ac{SNR}s. Audio examples of all the methods can be found online \footnote{\url{https://vikastokala.github.io/bccrn/}}.


\section{Conclusion}
\label{sec:conclusion}
In this paper, an end-to-end binaural speech enhancement method using a complex convolutional recurrent network is proposed. A loss function that optimizes the network for noise reduction, speech intelligibility, and human perception-based interaural cue preservation is proposed. The results of the experiments indicate that the suggested technique successfully reduced noise while preserving \ac{ILD} and \ac{IPD} information in the enhanced output. Additionally, the proposed method yielded better estimated binaural speech intelligibility compared to the baseline methods.

\bibliographystyle{IEEEtran}
\bibliography{sapstrings,sapref}
%
%
%
%
%
%
%
%

\end{sloppy}
\end{document}